\documentclass[twocolumn]{article}

\usepackage{graphicx,amsmath}

\newcommand{\Heph}{\hat{H}_{\mbox{qp-ph}}}
\newcommand{\He}{\hat{H}_{\mbox{qp}}}
\newcommand{\Hph}{{H}_{\mbox{ph}}}
\newcommand{\Hpho}{{H}_{\mbox{ph}}^{\mbox{os}}}
\newcommand{\Hphc}{{H}_{\mbox{ph}}^{\mbox{co}}}
\newcommand{\Hphh}{{H}^{\mbox{harm}}_{\mbox{ph}}}
\newcommand{\Hphoh}{{H}_{\mbox{ph}}^{\mbox{os-harm}}}
\newcommand{\Hphch}{{H}_{\mbox{ph}}^{\mbox{co-harm}}}

\begin{document}

\title{Dynamical two electron states in a Hubbard-Davydov model}

\author{L. Cruzeiro-Hansson$\dagger$, J.C. Eilbeck, J.L. Mar\'{\i}n and
  F.M. Russell\\ 
Department of Mathematics, Heriot-Watt
  University, Edinburgh
  EH14 4AS, U.K.\\
  $\dagger$ current address: CCMAR and FCT, Universidade do Algarve, Campus de Gambelas,\\
  8000 Faro, Portugal. }
\date{\today}

\maketitle
\begin{abstract}
  We study a model in which a Hubbard Hamiltonian is coupled to the
  dispersive phonons in a classical {\em nonlinear} lattice.  Our
  calculations are restricted to the case where we have only two
  quasi-particles of opposite spins, and we investigate the dynamics
  when the second quasi-particle is added to a state corresponding to
  a minimal energy single quasi-particle state.  Depending on the
  parameter values, we find a number of interesting regimes.  In many
  of these, discrete breathers (DBs) play a prominent role with a localized
  lattice mode coupled to the quasiparticles.  Simulations with a
  purely {\em harmonic} lattice show much weaker localization effects.  Our
  results support the possibility that DBs are important in HTSC.
\end{abstract}


\section{Introduction.}

In spite of the many studies \cite{hal88,am94,s95,mih95,d96} made
since it was first discovered \cite{bed86}, high temperature
superconductivity (HTSC) remains a challenge. The nature of the
carriers and the mechanism behind pair formation are still unclear.
According to many researchers, HTSC can be explained by a purely
electronic model, such as that described by the $t-J$ or the Hubbard
Hamiltonians, for which charge and/or spin interactions are paramount.
This view is essentially based on the absence of isotope effects seen
in some experiments \cite{wil98} and the apparent $d$-symmetry of the
superconducting wavefunction.  However, accumulating experimental
evidence exists for electron-lattice effects in high temperature
superconductors \cite{mor91,tmjc00,lan01}, and theories based on
electron-phonon interactions have also been proposed
\cite{am94,s95,d96,mih95}. Here we follow the idea that both
electronic correlations and electron phonon interactions are important
\cite{bh01} and study a model in which a Hubbard Hamiltonian is
coupled to dispersive phonons. Our main aim is to explore one extra
ingredient, which has generally been ignored until now, the importance
of the anharmonic character of lattice vibrations.  Whilst our
ultimate aim is to understand HTSC, here we propose a specific
mechanism for pair formation that involves the interaction of polarons
through a nonlinear lattice mode, which will have applications in
other areas. We study the stability of such a pair as a function of
the electron-electron (or hole-hole) interaction.

\section{The Hubbard-Davydov Hamiltonian.}
The Hamiltonian $\hat{H}$ we use has three parts:
\begin{equation}
\hat{H}= \He + \Heph + \Hph \label{htot}
\end{equation}
where $\He$ is the Hamiltonian for a quasiparticle with spin $\frac12$,
$\Heph$ describes the interaction of the quasiparticle with the
lattice and $\Hph$ is the lattice (phonon) Hamiltonian.

The Hamiltonian for the quasiparticle is the 1D Hubbard Hamiltonian:
\begin{eqnarray}
  \He & = & \epsilon \sum_{n,\sigma} \left( \hat{c}^\dagger_{n \sigma}
      \hat{c}_{n \sigma} \right) + \gamma  \sum_{n}
    \hat{c}^\dagger_{n \uparrow} \hat{c}_{n \uparrow}\label{hex}
  \hat{c}^\dagger_{n \downarrow} \hat{c}_{n \downarrow} \\
    \nonumber & & - t \sum_{n,\sigma} \left( \hat{c}_{n \sigma}^\dagger
        \hat{c}_{n-1 \sigma } + \hat{c}_{n \sigma}^\dagger
        \hat{c}_{n+1 \sigma} \right)
\end{eqnarray}
where the sums are over the sites $n$, going from 1 to $N$, ($N$
is the total number of lattice sites) and $\sigma$ refers to the
spin and can be up or down. $\hat{c}^{\dagger}_{n \sigma}$ is the
creation operator for a quasiparticle of spin $\sigma$ at site n.
$\epsilon$ is the self-energy of the quasiparticle, $t$ the
transfer term for the quasiparticle to move between neighbouring
sites. We depart from the usual notation in that the on-site
quasiparticle-quasiparticle coupling is here designated as
$\gamma$ (and not $U$) to avoid confusion with the variables $\{
u_n \}$ used for lattice displacement (see below). Both negative
and positive values of $\gamma$ will be considered, corresponding
to the attractive and repulsive Hubbard models, respectively.

As in the Davydov model for energy transfer in proteins
\cite{Sco92}, $\Heph$, the Hamiltonian for the interaction of the
quasiparticle with the lattice includes the coupling to acoustic
(or Debye) phonons:
\begin{equation}
  \Heph = \chi \sum_{n, \sigma} \left[ \left( u_{n+1}-u_{n-1} \right)
      \left( \hat{c}^{\dagger}_{n \sigma} \hat{c}_{n \sigma} \right)
    \right] \label{hep}
\end{equation}
where $\chi$ is a parameter which describes the strength of the
quasiparticle-lattice interaction. Many previous publications have
included electron-phonon interactions in the framework of the model of
Holstein \cite{hols59b}, in which only short-range interactions are
are considered. As has been pointed out elsewhere \cite{ak02}, when
the electron screening is poor, such as in cuprates, electron-phonon
interactions are long range, which can be described by acoustic
phonons.

The phonon Hamiltonian is as follows:
\begin{align}
  \Hph & =  \Hphc + \Hpho \label{hph}\\
  \nonumber \Hphc & = \frac{\kappa a^2}{72}\sum_{n=1}^{N} \left[
    \left(\frac{a}{a+u_n-u_{n-1}}\right)^{12} -\right.\\
  \nonumber &  \left. \qquad \qquad\qquad 2
  \left( \frac{a}{a+u_n-u_{n-1}} \right)^{6} \right]  \\
  \nonumber \Hpho & = \kappa^{\prime} \sum_{n=1}^{N} \left(
    \frac{1}{2} u_{n}^2 + \frac{1}{4} u_n^4 \right) + \frac{1}{2 M}
  \sum_{n=1}^N p_n^2
\end{align}
where $u_n$ is the displacement from equilibrium position of site $n$,
$p_n$ is the momentum of site $n$, $a$ is the equilibrium distance
between sites, $\kappa$ is the elasticity of the nonlinear lattice and
$\kappa^{\prime}$ is a similar constant for the on-site potential.
Here, the coupling interactions between sites are described by a
Lennard-Jones potential, $\Hphc$, a potential commonly used to
describe interactions between atoms.  In a high temperature cuprate,
this potential describes the interactions of the copper and oxygen
atoms in one Cu-O layer.  The on-site potential $\Hpho$ is as used in
many breather studies \cite{FW98}. It can be considered to represent
the effect, in a mean field approach, of the rest of the crystal on
the one dimensional chain whose states are studied explicitly. In a
cuprate, this models the effect of the neighbouring layers on the
Cu-O layer.

Our Hamiltonian (\ref{htot}-\ref{hph}) includes two sources of
nonlinear effects.  The first comes from the intrinsic nonlinearity
of the Lennard-Jones potential, $\Hphc$ and the on-site
potential, $\Hpho$. The second source of nonlinearity is
extrinsic and comes from the interaction of the quasiparticle with
the lattice (cf. Eq.~\ref{hep}).  The former is the source of
nonlinearity in the studies of discrete breathers \cite{FW98} and
the latter is the cause of localization in polaron theory.

We adopt a mixed quantum-classical approach in which the lattice is
treated classically, while the quasiparticle is treated quantum
mechanically. Accordingly, the displacements $u_n$ and momenta $p_n$
are real variables. The quasiparticle variables are operators, a
distinction which is marked by the hats above the operators. The
importance of quantum effects of the lattice can be assessed by
considering the full quantum model at finite temperature, which has
already been done for the Davydov Hamiltonian. It was found that, at
0.7 K, the lattice displacement correlated with the position of the
quantum particle in exact semiclassical Monte Carlo simulations
differed by 15 \% from the corresponding variable in exact simulations
in the fully quantum system. At 11.2 K, the two approximations lead to
virtually the same value \cite{chk95}. We would like to emphasize that
the approximation we consider here is {\em not} an adiabatic
approximation. In an adiabatic approximation the kinetic energy of the
phonons is neglected with respect to the kinetic energy of the quantum
particle. We do not do that here, as our dynamical equations, eqs (6)
and (7) below, include the time derivative of the momenta of the
lattice sites. What we do is to consider that the dynamics of the
lattice can be treated classically. Both the semiclassical (or
quantum/classical, as we prefer to call it to differentiate from other
use in the literature) approach we apply here and the adiabatic
approximation lead to similar results when we consider the ground
states of the system (because the corresponding solutions have zero
kinetic energy), but they are different when we deal with dynamics, as
we do in this work.

Ultimately, the need for a full quantum treatment comes from
comparison with experimental results. Isotopic effects can only be
described in a fully quantum framework. Our main aim here is to
explore the importance of anharmonicity in the lattice for the
dynamics of paired states, something which is much more complicated to
do within a fully quantum formalism. Thus, as a first approximation,
we restrict ourselves to the mixed quantum-classical regime and study
the behaviour of a pair of quasiparticles, coupled to a nonlinear
lattice.

With these assumptions, the exact two quasiparticle wavefunction for
the Hamiltonian (\ref{htot}-\ref{hph}) is:
\begin{equation}
| \psi (t) \rangle = \sum_{n,m=1,N} \phi_{nm} (\{ u_n \}, \{ p_n \},t)
               \, \hat{c}^{\dagger}_{n \uparrow} \,
               \hat{c}^{\dagger}_{m \downarrow} |0\rangle
\label{wf}
\end{equation}
where $\phi_{nm}$ is the probability amplitude for a quasiparticle
with spin up to be at site $n$ and a quasiparticle with spin down to
be at site $m$.  The probability amplitude is dependent on the lattice
displacements and momenta in a way that is not specified {\it a
  priori} and is determined by the equations of motion.  Similarly to
other systems \cite{CT97}, the equations of motion for probability
amplitudes $\phi_{nm}$ are derived by inserting the wavefunction
(\ref{wf}) in the Schr\"odinger equation for the Hamiltonian
(\ref{hex}-\ref{hph}), and the equations for the displacements and
momenta are derived from the Hamilton equations for the classical
functional ${\cal E}^2=\langle\psi | \hat{H} | \psi\rangle $.  They
are:
\begin{align}
  \imath \hbar \frac{ d \phi_{jl} }{d t} & =  -t \left( \phi_{j-1 l}
    + \phi_{j+1 l} + \phi_{j l-1} + \phi_{j l+1} \right) + \gamma
  \phi_{jl} \delta_{jl} + \nonumber\\  & \qquad +\chi \left(u_{j+1} -
    u_{j-1} + u_{l+1} - u_{l-1} \right) \phi_{jl} \label{eqphi}
  \\
  \frac{ d p_j}{d t} & =  - \frac{\partial \Hph}{\partial
    u_j}\label{eqp} \\ \nonumber & \quad \quad 
-\chi
  \left( |\varphi^{\uparrow}_{j-1}|^2 - |\varphi^{\uparrow}_{j+1}|^2 +
    |\varphi^{\downarrow}_{j-1}|^2 - |\varphi^{\downarrow}_{j+1}|^2
  \right)
\end{align}
where $|\varphi^{\uparrow}_j|^2$, the probability for the
quasiparticle with spin up to be in site j and
$|\varphi^{\downarrow}_j|^2$, the probability for the
quasiparticle with spin down to be in the same site. These are
given by:
\begin{align*}
  |\varphi^{\uparrow}_j|^2  &=   \langle \psi | \hat{c}^{\dagger}_{j
    \uparrow} \hat{c}_{j \uparrow}| \psi\rangle  =\sum_{l=1}^N
  |\phi_{jl}|^2, \\
  |\varphi^{\downarrow}_j|^2  &=  \langle \psi |
  \hat{c}^{\dagger}_{j \downarrow} \hat{c}_{j \downarrow}|
  \psi\rangle =\sum_{l=1}^N |\phi_{lj}|^2 
\end{align*}

\section{Dynamical states.}

We consider the case in which the quasiparticle density is low and
the starting point is that of an isolated quasiparticle
interacting with the lattice. We wish to find if the addition
of a second quasiparticle with opposite spin to that state can
lead to pairing of the two quasiparticles, and how the relative
stability of the paired state depends on the
quasiparticle-quasiparticle interaction $\gamma$.

We start from the state of a single quasiparticle. The
wavefunction is
\begin{equation}
|\psi^1_{\sigma}\rangle  = \sum_{n} \phi^1_n \hat{c}^{\dagger}_{n
|\sigma} |0\rangle
\end{equation}
Minimum energy states for this one quasiparticle can be found by
numerical minimization of the energy functional ${\cal E}^1 =
\langle\psi^1| \hat{H}| \psi^1\rangle $ with respect to the
probability amplitude for a single quasiparticle in site $n$,
$\phi^1_n$, and to the displacements $u_n$ \cite{CK94}. Two kinds
of minimum energy states are found. For sufficiently large
quasiparticle-lattice interaction $\chi$, the quasiparticle states
are localized and there is an associated lattice distortion. We
call this the single particle polaron, or simply polaron. Below a
threshold value for $\chi$, the states are delocalized, as in the
usual Bloch states, and the lattice is undistorted. We have
considered a value of $\chi$ and other parameters such that the
initial one quasiparticle polaron state is neither too weak nor
too stable when compared with delocalized, Bloch states for the
same values. While it is important to find the behaviour of the
two quasiparticle states considered here for different values of
the parameters, our choice ensures that the results here are not
the consequence of extreme values.

The dynamical states we study are the perturbations of the single
polaron state, induced by the presence of a second quasiparticle
with opposite spin. Because the number of variables $\phi_{nm}$
 that characterize the wavefunction (\ref{wf}) increases
with the square of the lattice size, in order to be able to
integrate the equations of motion for a sufficiently long time,
the size of the lattice was kept relatively short, i.\ e.\ the
number of sites is $N=20$. The aim is to investigate the influence
of the strength and sign of the quasiparticle-quasiparticle
interaction $\gamma$ on the dynamics of the paired quasiparticle
states.

\begin{figure} 
  \begin{center}
      \includegraphics{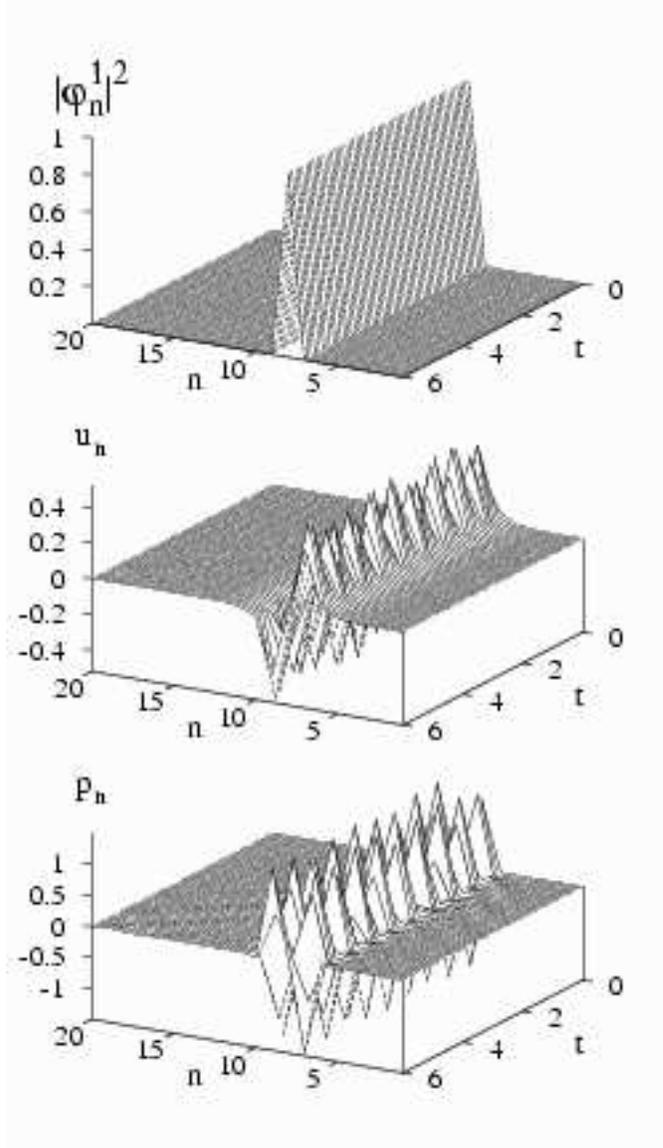}
  \end{center}
  \caption{Time dependence for (a) the probability for one
    quasiparticle to be in site $n$, ($n = 1 \cdots N$, $N=20$), (b)
    the lattice displacement and (c) the momentum of site $n$. Time is
    in picoseconds. The parameters are $t =10 \times 10^{-22}$J, $\chi =
    100$pN, $\kappa = 1 $N/m, $\kappa^{\prime}= 2 \kappa$, $a = 4.5
    \AA$ and $\gamma = -100 \times 10^{-22}$J.}
    \label{gm100}
\end{figure}

\begin{figure} 
  \begin{center}
    \includegraphics{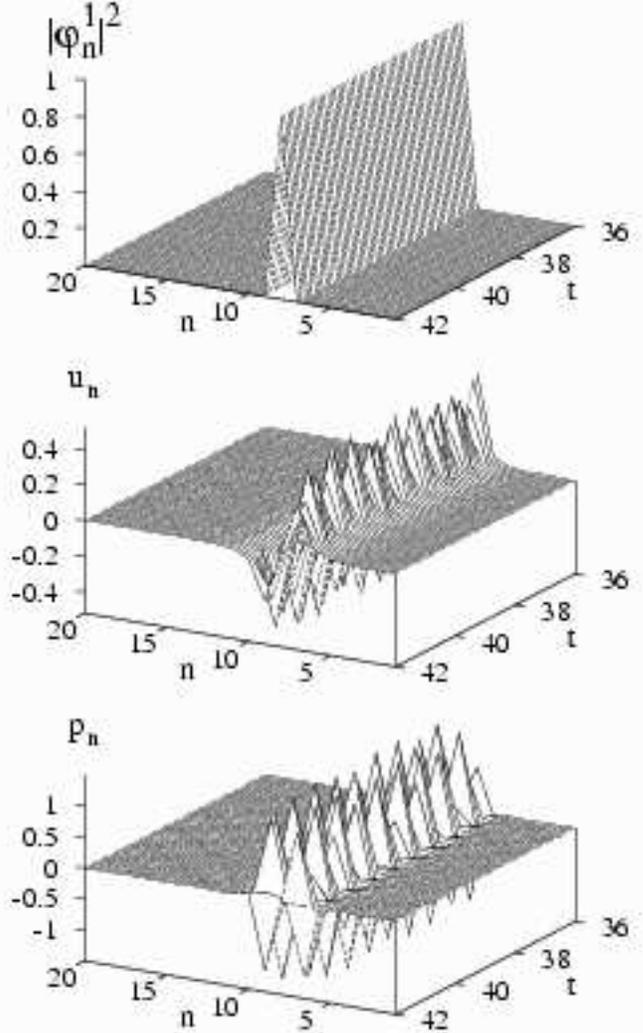}
  \end{center}
  \caption{Same as Fig.~\ref{gm100}, at a later time.}
    \label{gm100c}
\end{figure}

The parameters of the simulations in the figures are the same, except
for the quasiparticle-quasiparticle interaction $\gamma$.
In Fig.~\ref{gm100} we set $\gamma/t=-10$ in an {\em attractive} Hubbard
model.
The addition of a second quasiparticle leads to a localized
state for the pair, with a very slight peak oscillation, that is
hardly visible in the figure.  (The probability for the second
quasiparticle is the same as that shown and is not displayed). The
lattice, however, sets into a breather-like oscillation \cite{FW98},
i.e., a localized excitation with an internal oscillation. Indeed, at
the site of the initial lattice distortion, oscillations are clearly
visible in the lattice displacements and momenta.  A striking
observation is that the amount of radiation generated is very small,
and most of the energy of the lattice is associated with the breather.
Fig.~\ref{gm100c}, which displays another 6 ps period of the dynamics
at a later time, demonstrates the stability of this solution.

\begin{figure} 
  \begin{center}
    \includegraphics{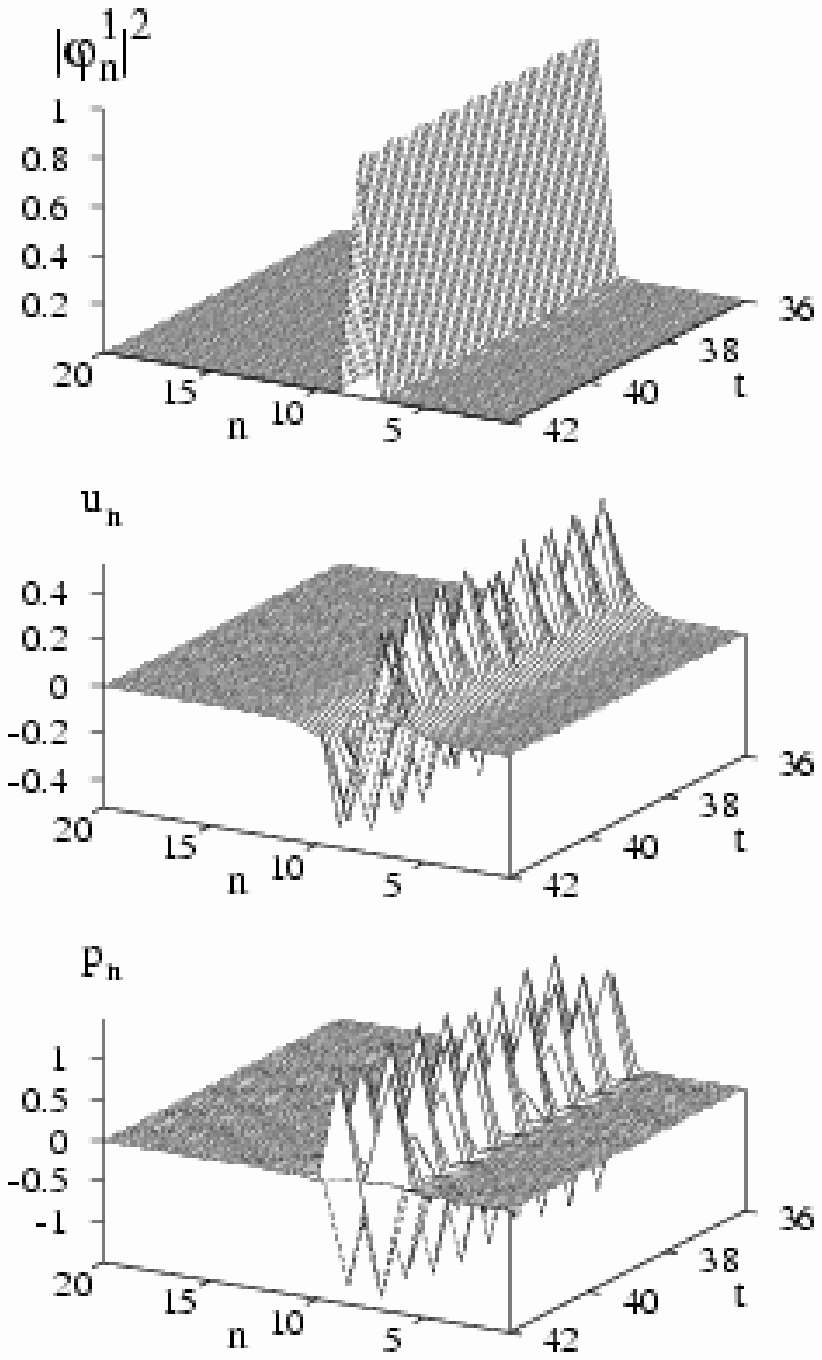}
\end{center}
  \caption{Same as Fig.~\ref{gm100}, but with $\gamma= - 5 \times
    10^{-22}$J.}
\label{gm5c}
\end{figure}

A Hubbard Hamiltonian with a much weaker attraction, corresponding to
a ratio of $\gamma /t = - 0.5$, is considered in Fig.~\ref{gm5c},
where the last 6 picoseconds of a 42 picosecond simulation are
displayed.
A modulation of the peak of the probability distribution is
now clearly seen, which has the same frequency as the main modulation
of the lattice breather. The modulation of the quasiparticle
probability is associated with a periodic change of shape in which a
lower peak with a slight tail appears. Even at this comparatively much
weaker interaction, the amount of radiation is very small and most of
the lattice energy is in the breather. The frequency of the main
modulation of the breather is as for $\gamma/t = -10$.

\begin{figure} 
  \begin{center}
    \includegraphics{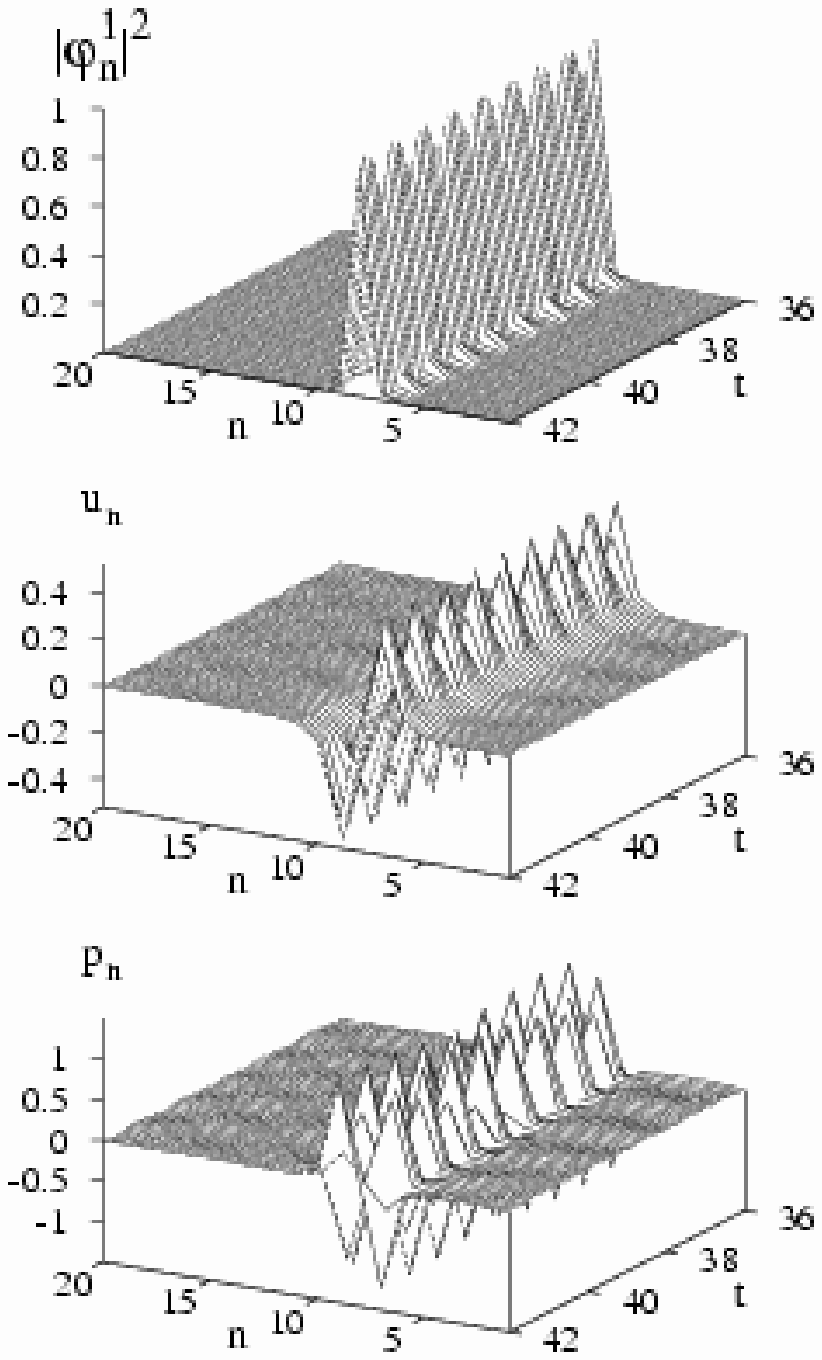}
\end{center}
  \caption{Same as Fig.~\ref{gm100}, but with $\gamma= + 5 \times
    10^{-22}$J.}
\label{g5c}
\end{figure}

In Fig.~\ref{g5c} a {\em repulsive} Hubbard Hamiltonian is considered,
with $\gamma / t = + 0.5$.
The modulations and the associated tails of the probability
distribution for the quasiparticle are now more pronounced, but their
main frequency is unchanged. Although there is a slight increase in
the radiation in the lattice, the stability of the breather and of the
quasiparticle solution is apparent.

\begin{figure} 
  \begin{center}
    \includegraphics{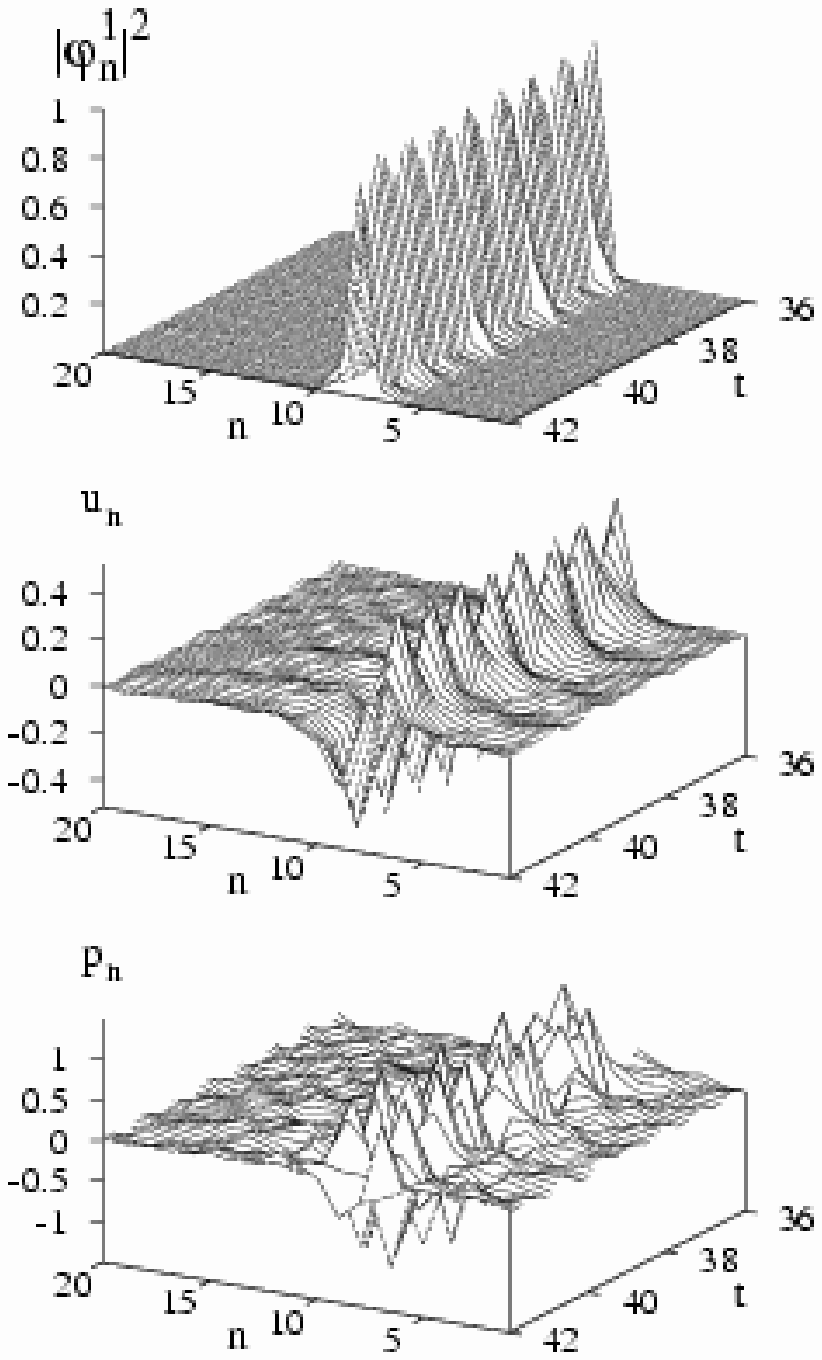}
\end{center}
  \caption{Same as Fig.~\ref{gm100}, but with $\gamma= + 10 \times
    10^{-22}$J.}
\label{g10c}
\end{figure}

In Fig.~\ref{g10c} the repulsive interaction is increased to $\gamma/t
= + 1$.
The modulations in the probability distribution for the
quasiparticles lead to greater periodic changes of shape, still with
the same frequency as for the other values of $\gamma$. The radiation
in the lattice is now more visible, but the breather remains stable.

\begin{figure} 
  \begin{center}
    \includegraphics{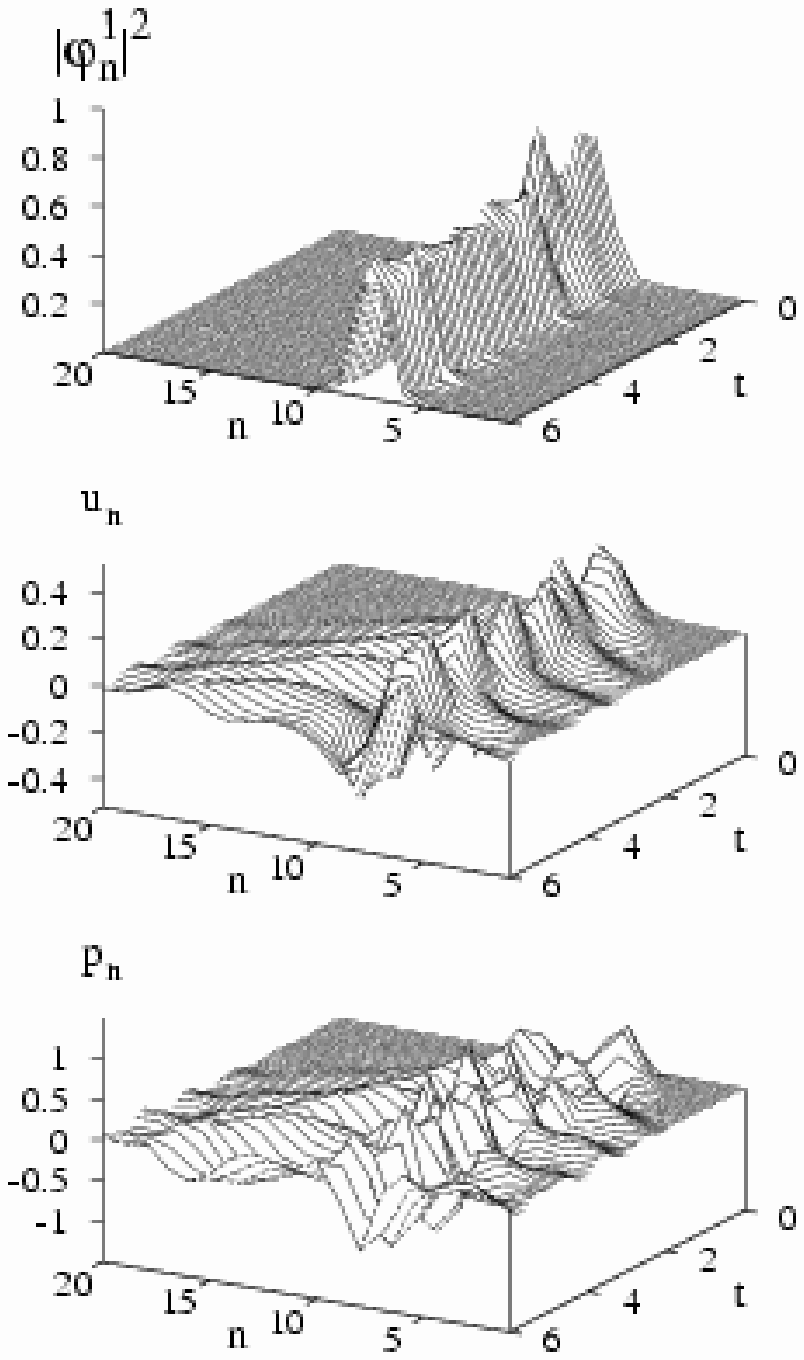}
\end{center}
  \caption{Same as Fig.~\ref{gm100}, but with $\gamma= + 50 \times
    10^{-22}$J.}
\label{g50}
\end{figure}

\begin{figure} 
  \begin{center}
    \includegraphics{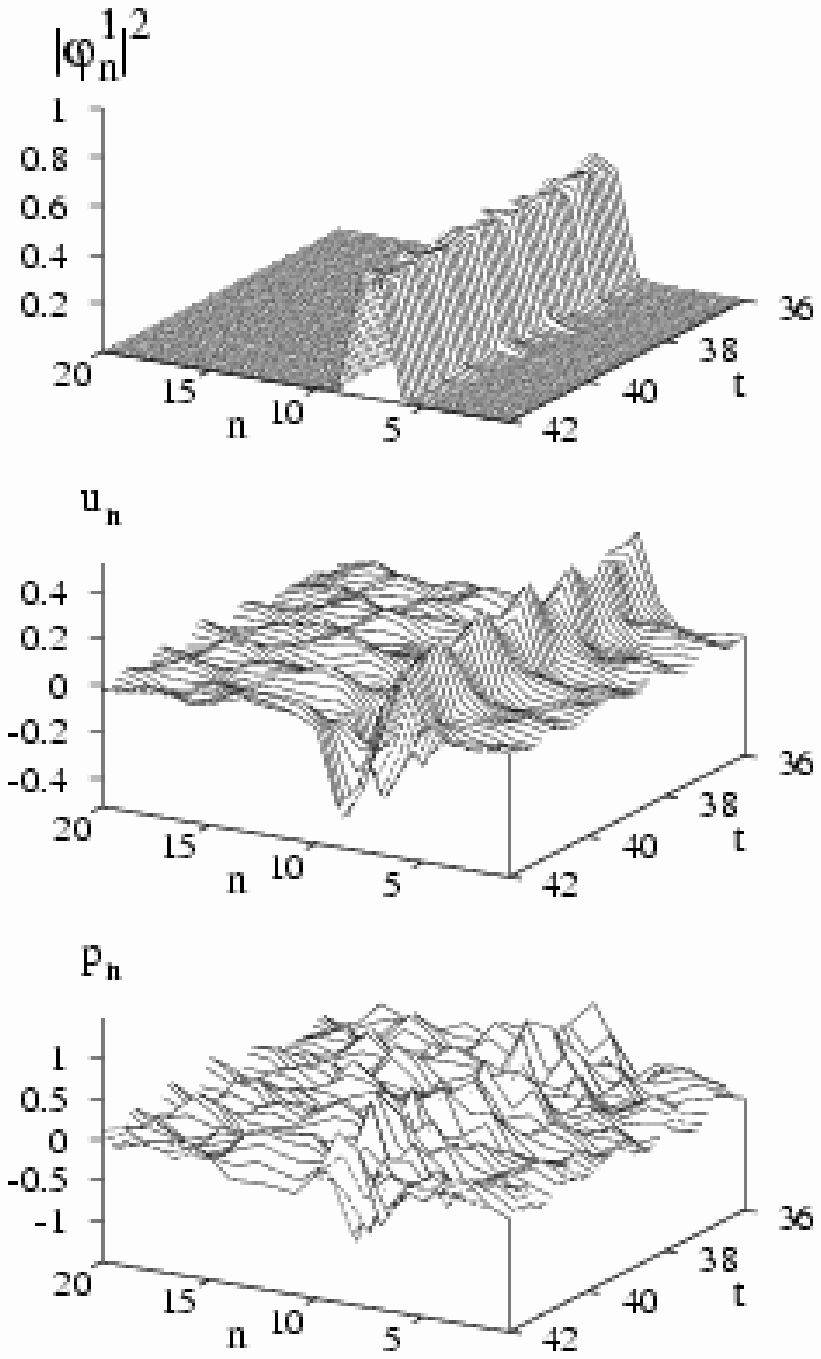}
\end{center}
  \caption{Same as Fig.~\ref{g50}, but at a later time.}
\label{g50c}
\end{figure}

In Figures \ref{g50} and \ref{g50c}, a large repulsive value,
corresponding to $\gamma / t = 5$ is taken.
This leads to a change in the probability distribution for the
quasiparticles, from a single site peak into a two site peak, with
periodic oscillations which make one probability at one site larger
than the other. The lattice variables show that, concurrently with the
appearance of the breather, a considerable amount of radiation is
generated.  Also noticeable is the fact that the frequency of the
modulations has changed. Fig.~\ref{g50c} shows that the new
quasiparticle probability distribution is stable, as well as the
lattice breather, even if the noise which results from successive
passes of the radiation through the periodic boundaries, constitutes a
significant part of the lattice energy.

\begin{figure} 
  \begin{center}
    \includegraphics{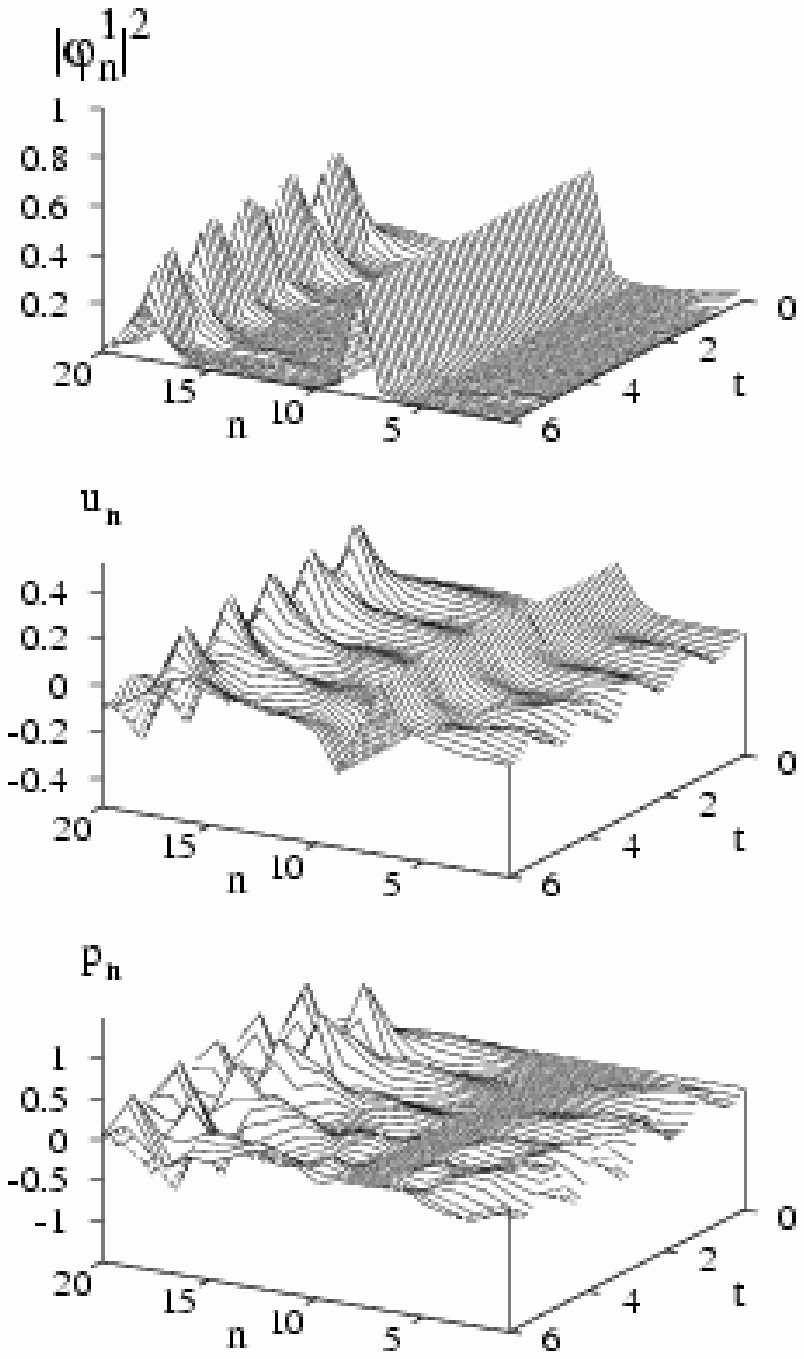}
\end{center}
  \caption{Same as Fig.~\ref{gm100}, but with $\gamma= + 100 \times
    10^{-22}$J.}
\label{g100}
\end{figure}

\begin{figure}
  \begin{center}
    \includegraphics{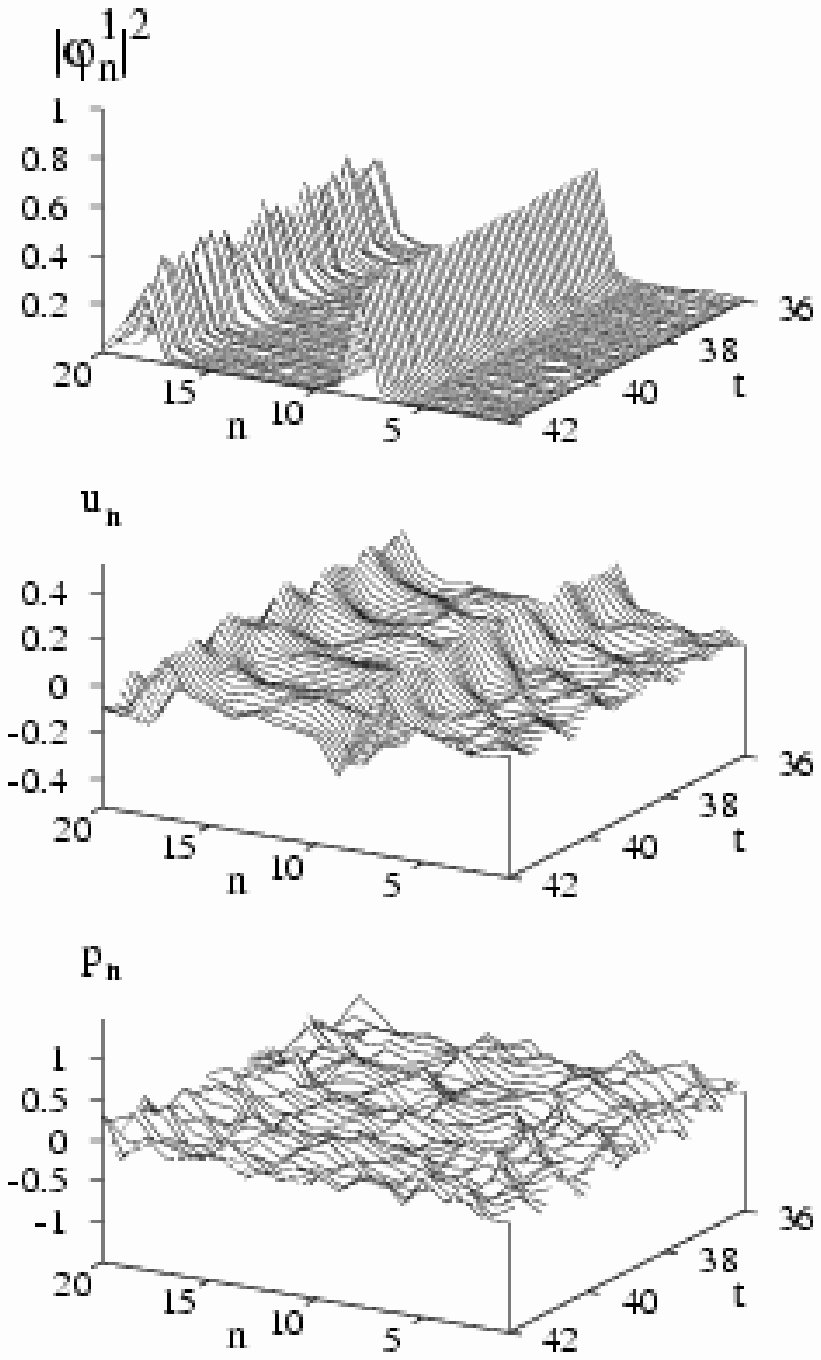}
\end{center}
  \caption{Same as Fig.~\ref{g50}, but at a later time.}
\label{g100c}
\end{figure}

In Figures \ref{g100} and \ref{g100c}, a repulsive interaction
corresponding to $\gamma / t = 10$ is used.
 Fig.~\ref{g100} shows
that a drastic transformation takes place in which the initial
distribution changes into a two peak distribution. One of the
peaks is located where the initial lattice distortion was and the
second peak is as far away from it as it can be in this lattice.
Also, while the peak that is located at the original lattice
distortion site remains unmodulated in time, as well as its
associated lattice distortion, the second peak oscillates with
approximately the same frequency as that in Figs. \ref{g50} and
\ref{g50c}. The momenta in Fig.~\ref{g100} show clearly that the
second peak has an associated lattice breather, while the first
peak is associated with a distortion that is essentially static.
After some time, because of the repeated reflection of the
radiation from the boundaries, this picture is not so clear. Both
peaks show oscillations in the displacements and the momenta of
the lattice are rather noisy. However, Fig.~\ref{g100c} does
illustrate the stability of the two peak solution, even in the
presence of such relatively large amount of noise.

\section{Dynamical states in the fully harmonic approximation}

The early theory of pair formation via interaction with phonons
assumed that the lattice motion was harmonic. It is interesting to see
how the dynamics of the two electron states would be in this case, and
this section is devoted to that question.  The first two terms
(\ref{hex}), (\ref{hep}) in the Hamiltonian we consider in this section
are the same as before, but now the phonon Hamiltonian is given by:
\begin{align}
  \Hphh & =  \Hphch + \Hphoh \label{harm}\\ \nonumber
  \Hphch & =  \frac{1}{2} \, \kappa \sum_{n=1}^{N} \left( u_n - u_{n-1}
  \right)^2, \\ \nonumber
  \Hphoh   &=   \kappa^{\prime} \sum_{n=1}^{N} \left( \frac{1}{2} u_{n}^2
   \right) + \frac{1}{2 M} \sum_{n=1}^N p_n^2
\end{align}

The phonon Hamiltonian (\ref{harm}) can be obtained from
(\ref{hph}) by considering the limit of small displacements, in
which only the linear terms remain. In this case, the only
nonlinear term left for the total Hamiltonian is that which
describes the quasiparticle-lattice interaction. It should be
pointed out that, if we disregard the correlation term in
(\ref{hex}), the equations of motion for this system are those
studied by a number of authors \cite{hols59b,km93} for a single
single polaron, and for any number of polarons by Alexandrov
\cite{a01}.

\begin{figure}
  \begin{center}
       \includegraphics{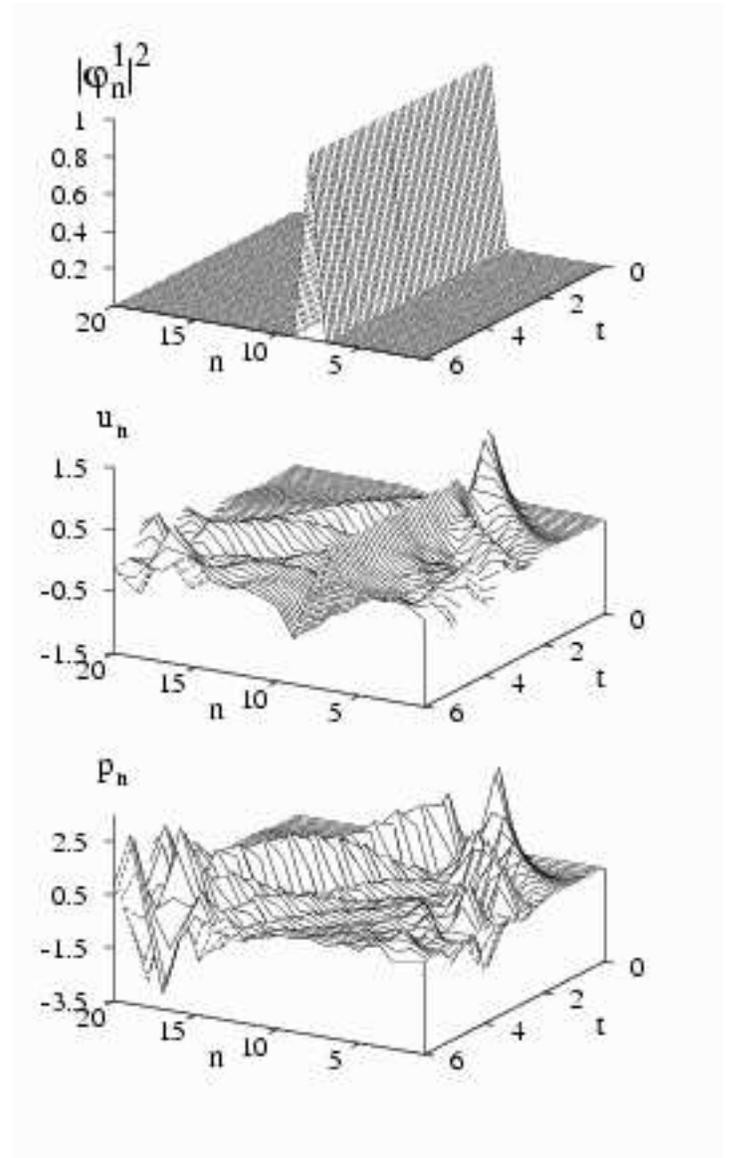}
\end{center}
  \caption{Same as Fig.~\ref{gm100}, but with $\gamma= - 100 \times
    10^{-22}$J and for the harmonic lattice \ref{harm}.}
\label{hgm100}
\end{figure}

Fig.~\ref{hgm100} shows that when the effective interaction is
such that $\gamma / t = -10$, the addition of an extra electron to
the minimum energy single polaron leads to a state in which both
electrons are in the same site with a strong lattice deformation
of breather type associated with their presence.
 The time
evolution of the momenta, however, shows that there is no breather
formation, only phonons which travel along the lattice. Because of
the periodic boundary conditions, these phonons eventually come
back and after they have crossed each other many times the lattice
becomes very noisy.  The lattice deformation associated with the
two electrons oscillates periodically because of the interference
of these phonons, but does not move. Also, the state of the two
electrons remains localized on one site all the time.

Similar dynamics takes place for $\gamma / t = -0.5$, except
that very slight oscillations in the probability distribution for
the electron states also takes place (not shown).

\begin{figure}
  \begin{center}
 \includegraphics{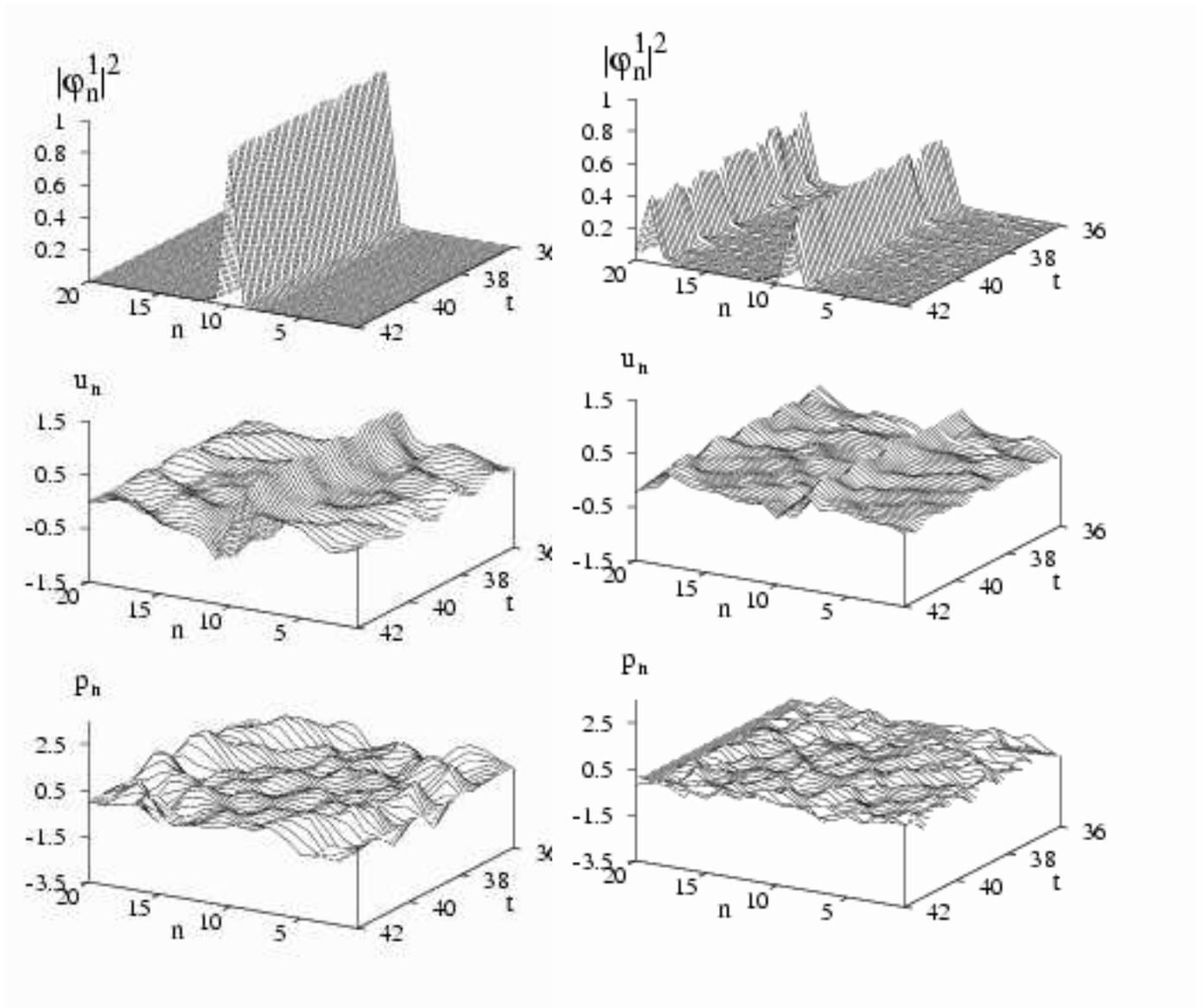}
\end{center}
  \caption{Same as Fig.~\ref{gm100}, but with $\gamma= + 50 \times
    10^{-22}$J and for the harmonic lattice \ref{harm}.}
\label{hg50}
\end{figure}

When the electron-electron interaction is repulsive and such that
$\gamma /t = +5$, the phonon emission leads to fluctuations in the
electron probability distribution that are clearly visible in
Fig.~\ref{hg50}.
 The dynamics is similar to that of
fig.~\ref{hgm100}, with phonons propagating along the lattice and
causing oscillations in the otherwise constant distortion induced
by the two electrons. Again, the momenta show that there is no
breather formation and all the dynamics of the lattice is due to
the phonon propagation and interference.

\begin{figure}
  \begin{center}
  \includegraphics{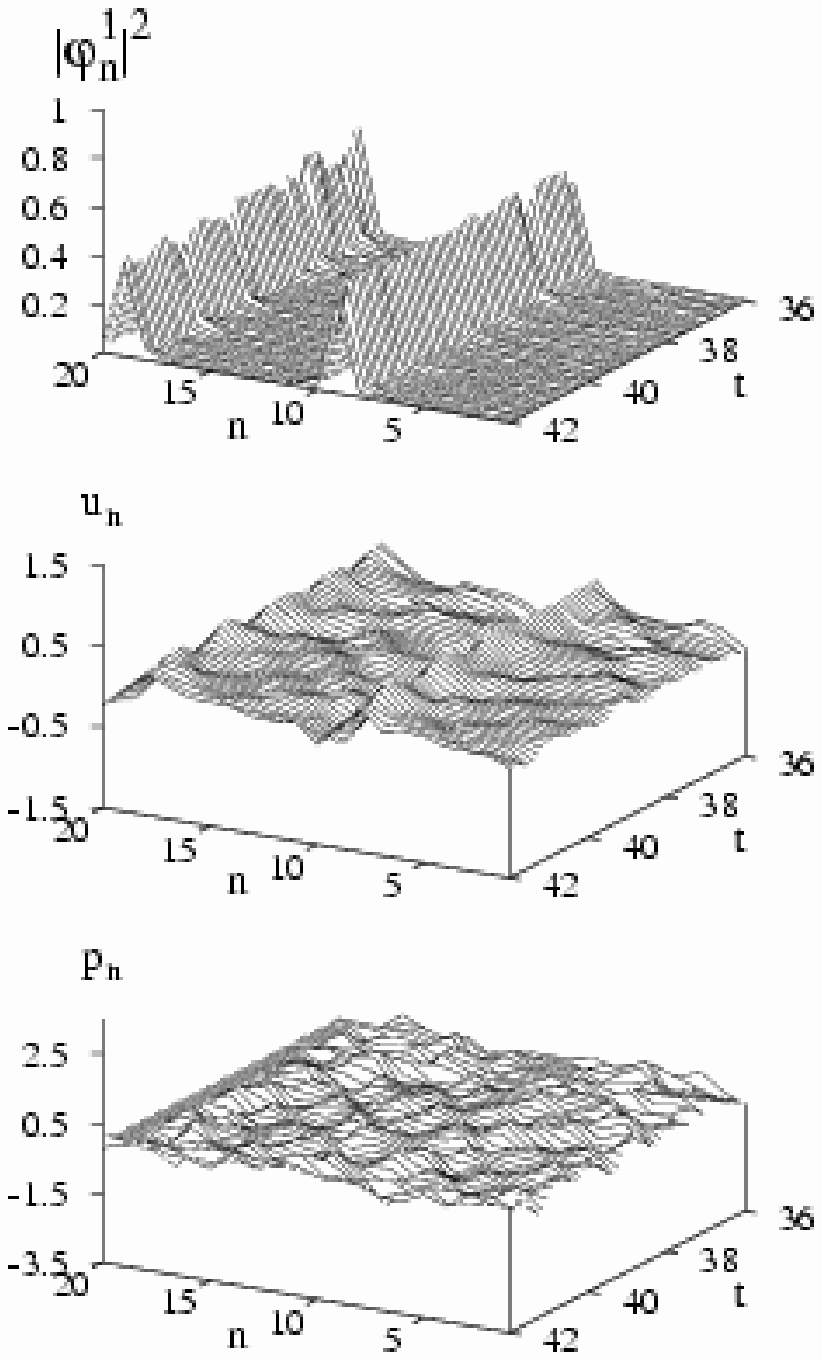}
\end{center}
  \caption{Same as Fig.~\ref{gm100}, but with $\gamma= + 100 \times
    10^{-22}$J and for the harmonic lattice \ref{harm}.}
\label{hg100}
\end{figure}

For a repulsive interaction for which $\gamma / t = +10$, the two
electrons split up and the probability distribution shows two
peaks, both of which have an associated lattice deformation with
the breather profile (see Fig.~\ref{hg100}).
Phonons are generated from each of these locations and their
interference eventually leads to a noisy lattice. The two peaks in the
probability distribution for the electrons oscillate in a less regular
fashion than in the anharmonic lattice, but remain stable throughout
the simulation. It should be noticed that for this harmonic
approximation also, the lattice displacements induced by the
electrons/holes are not small.  Hence, an accurate representation of
the dynamics should include the nonlinear terms in the lattice
Hamiltonian, as was done in the previous section.

\section{Discussion}

Our aim was to investigate the relative stability of a correlated
pair of quantum quasiparticles with opposite spins with respect to
their uncorrelated states. The starting point was thus the state
of a single quasiparticle polaron and we studied the dynamic
states which arise when a second quasiparticle is added to the
first state. The Hamiltonian used includes several physical
ingredients. On the one hand, it contains two sources of
nonlinearity, one intrinsic to the lattice and another which
arises from the quasiparticle lattice interaction. Such nonlinear
lattices have been shown to possess generic solutions known as
discrete breathers (DBs) \cite{FW98}. The study of systems in
which nonlinear lattices are coupled to one quantum quasiparticle,
on the other hand, is just beginning \cite{au97,cemr00}.

To our knowledge, this is the first time that the coupling of two
quantum quasiparticles to a nonlinear lattice has been considered.
Indeed, a second ingredient is the inclusion of
quasiparticle-quasiparticle interactions, in addition to the
quasiparticle-lattice interactions found in the polaron model.
The quasiparticle-quasiparticle interactions can represent Coulomb
interactions, and/or spin-spin interactions, and be either
attractive or repulsive. The dynamical simulations indicate that
for these extended systems, DBs are generic solutions also
and can be generated by the presence of a second quasiparticle.
These lattice breathers can in turn stabilise localized, paired,
quasiparticle states, for a large range of $\gamma$ values.
Windows of $\gamma$ were found for which similar solutions are
obtained. Thus, for a ratio of $\gamma/t$ between $-10$ and $+1$
(Figs.~\ref{gm100}-\ref{g10c}), DBs are found in the lattice and
in the quasiparticle, with the same main modulation frequencies.
For larger values of $\gamma/t$, two different solutions were
found (see Figs.~\ref{g50}-\ref{g100c}). In one solution the
quasiparticles distribution is split into equal values in two
neighbouring sites and in the second a two peak distribution, with
the peaks as far apart as possible in the lattice used, is
observed.

This Hamiltonian includes the two main physical causes for
quasiparticle pairing that have been considered in HTSC and allows
for interpolation between them, by varying the strength of the
relevant parameters. According to our results, a greater
importance of quasiparticle-lattice interactions in pair formation
should arise in systems for which the dynamics of the lattice is
fast enough compared to the quasiparticle dynamics, so that the
lattice relaxes when the two quasiparticles meet. Conversely, a
corresponding greater importance of quasiparticle-quasiparticle
interactions should be associated with systems in which the
lattice dynamics is much slower than the quasiparticle dynamics.

An implicit assumption in this study is that the nonlinear
character of the lattice plays an important role in HTSC. Although
the lattice distortions are weak in conventional superconductors,
and thus the lattice dynamics can be approximately described by a
linear system, we argue that in HTSC these distortions are such
that the lattice enters a nonlinear regime. This may be why the
sound velocity decreases by a few parts per million in
conventional superconductors, whereas in a high Tc material there
is an increase which is two or three orders of magnitude larger
than in the former case. Our simulations with the harmonic lattice
show that the percentage of energy transferred to travelling
phonons is much larger than for the anharmonic lattice.

The breather-like solutions found in the dynamical simulations are
a signature of the nonlinear dynamics of the lattice. The
possibility that breathers are associated with HTSC has been
suggested elsewhere \cite{rc96,mre01}. Our study indicates that DBs
are generic excitations in systems governed by the Hamiltonian
used here. Moreover, within a certain range of the parameters,
the states in which two quasiparticles are paired and coupled to a
DB are energetically more favourable than those of uncorrelated
quasiparticles. Hence, this study gives weight to the
possibility that DBs are important in HTSC.

\section*{Acknowledgments} JLM acknowledges a {\it Marie Curie} TMR
  fellowship from the EU (no.\ ERBFMBICT972761).  LC-H and JCE are
  grateful for travel support of a British Council Treaty of Windsor
  grant.  JCE and JLM would also like to acknowledge support from the
  EU under the LOCNET Research Training Network HPRN-CT-1999-00163.

\end{document}